\documentclass[a4paper,UKenglish,cleveref, autoref, thm-restate, numberwithinsect]{lipics-v2021}

\usepackage{cite}                       
\usepackage{amsmath,amsfonts,amssymb,amsthm}
\usepackage{graphicx}
\usepackage{algorithm}
\usepackage[noend]{algpseudocode}
\usepackage{float}
\usepackage{subcaption}
\usepackage{tikz}
  \usetikzlibrary{decorations.pathreplacing,positioning}
\usepackage{multirow}
\usepackage{textcomp,xcolor}
\usepackage{balance}                    

\title{Sublinear Edge Fault-Tolerant Hyperspanners for Hypergraphs} 

\titlerunning{Sublinear Edge Fault-Tolerant Hyperspanners for Hypergraphs} 

\author{Jialin He}{Colby College, USA}{}{}{}
\author{Nicholas Popescu}{Duke University, USA}{}{}{}
\author{Chunjiang Zhu\footnote{Corresponding author: czhu@odu.edu}}{Old Dominion University, USA}{}{}{}

\authorrunning{He et al.} 

\Copyright{He et al.} 



\category{} 

\relatedversion{} 

\nolinenumbers 

\EventEditors{}
\EventNoEds{0}
\EventLongTitle{}
\EventShortTitle{}
\EventAcronym{}
\EventYear{}
\EventDate{}
\EventLocation{}
\EventLogo{}
\SeriesVolume{}
\ArticleNo{}

\begin{document}

\maketitle

\begin{abstract}
In this paper, we initiate the study on fault-tolerant (FT) graph spanners for hypergraphs and show the generalization to hypergraphs in the FT setting is non-trivial. An FT spanner approximates shortest distances under network failures, widely used in applications such as routing and distributed computing. We first provide a systematic study on extending spanners to hyperspanners in both non-faulty and FT settings and reveal that the latter case is more interesting: simple methods can only produce a linear size in the number of allowed faults $f$, while all known optimal sizes of FT graph spanners are sublinear in $f$. Inspired by the FT clustering technique in Parter's paper \cite{partervft}, we propose a hypergraph clustering based algorithm with an improved sublinear size bound. Specifically, for an $n$-node $m$-edge hypergraph with rank $r$ and a stretch parameter $k$, our algorithm constructs edge FT (EFT) hyperspanners of stretch $2k-1$ and size $O(k(k+r)f^{1-1/(rk)}n^{1+1/k}\log n)$ with high probability in time $\widetilde{O}(mr^3+nrf)$ ($\widetilde{O}$ hides polylogarithmic factors). We also establish size lower bounds, $\Omega((f/r)^{r-1-1/k+o(1)}n^{1+1/k-o(1)})$ for vertex FT (VFT) hyperspanners and $\Omega(f^{1-1/r-1/(rk)+o(1)}n^{1+1/k-o(1)}+fn)$ for EFT hyperspanners, leaving a gap of $k(k+r)f^{1/r}$ yet to close. We believe that this work will spark interest in developing optimal-sized FT hyperspanners for hypergraphs.
\end{abstract}


\section{Introduction}

Graph spanners are sparse subgraphs that approximately preserve pairwise distances \cite{PS89}. A (\emph{multiplicative}) $\alpha$-spanner ensures that all shortest path distances are approximated by a factor of $\alpha$, called stretch factor, while an \emph{additive} $\beta$-spanner allows distances to increase by an additive surplus $\beta$. It was well-established that for an integer $k>1$, every undirected graph on $n$ vertices has a $(2k-1)$-spanner of size (number of edges) $O(n^{1+1/k})$ \cite{ADD+93,TZ05}. Graph spanners have found wide applications, such as routing \cite{peleg1989trade}, synchronizers \cite{awerbuch1990network}, distance oracles \cite{TZ05}, and preconditioning of linear systems \cite{EES+08}. While prior work focuses on simple graphs, higher order relationships in many real-world systems, such as biological pathways, social communities, and acquaintance networks, are more naturally modeled as \emph{hypergraphs}, where a hyperedge may contain more than two vertices. Extending spanner theory to hypergraphs is a natural and important direction. The scarcity of literature seems to suggest the extension is trivial: one can convert a hypergraph into an associated graph \cite{bansal2019new} by clique expansion (i.e., replacing each hyperedge by a clique on its vertices). Then any spanner in the associated graph can be translated to a hyperspanner in the hypergraph. See our detailed discussions in Appendix~\ref{apx:associated}.

However, the problem of constructing hyperspanners in the \emph{fault-tolerant (FT)} setting becomes non-trivial and more interesting. Many real-world networks are inherently prone to failures of vertices and edges, which well motivates FT graph structures. Informally, an $f$-edge-fault-tolerant (EFT) $t$-hyperspanner of a hypergraph $H$ is a sparse subhypergraph $S\subseteq H$ such that, after deleting any set $F$ of at most $f$ hyperedges, distances in $S\setminus F$ still approximate distances in $H\setminus F$ within factor $t$. Thus the spanner must preserve approximate distances not only in the original hypergraph, but also after every allowed failure event. The vertex-fault-tolerant (VFT) hyperspanners can be defined similarly. It is conceivable to apply the associated graph idea to construct VFT hyperspanners. This does not work unfortunately, since it is unclear how to set the number of faults. To see this, consider a hyperedge $h$ includes three vertices $v_1,v_2,v_3$ and it translates to a triangle in the associated graph. A vertex fault $v_1$ would remove the associated hyperedge $h$, which should remove the three edges of the triangle in the associated graph. However, in the associated graph the faulty vertex $v_1$ would not remove the non-associated edge $(v_2,v_3)$, inconsistent with the behavior in hypergraphs. 

The method of associated graphs is applicable to the construction of EFT hyperspanners though. Since a faulty hyperedge corresponds to at most $r^2$ faulty edges in the associated multi-graph (where $r$ is the rank of the hypergraph), the number of faults for the EFT spanner in the associated graph needs to be $fr^2$, instead of only $f$. As the size of EFT spanners in multi-graphs was settled to $\Theta(fn^{1+1/k})$ by \cite{petruschka2025color}, the size of EFT hyperspanners using this method is $O(fr^2n^{1+1/k})$. Indeed, we can reduce the size and remove the term $r^2$ by adapting the peel-off method of \cite{CLP+10} as we will elaborate later. Nonetheless, the size $O(fn^{1+1/k})$ is still linear in the number of faults $f$. Since all known optimal size of EFT and VFT spanners are sublinear in $f$ in simple graphs, a natural question is \emph{can we achieve a sublinearity in $f$ for FT hyperspanners?} In this work, we provide an affirmative answer for EFT hyperspanners accompanied with lower bounds on the size of VFT and EFT hyperspanners.

\begin{table*}[t]
    \centering
    \begin{tabular}{|c|c|c|}
    \hline
    Type & Spanner Size & Hyperspanner Size \\
    \hline
      \multirow{3}{*}{EFT} & $O(k^2f^{1/2-1/(2k)} n^{1+1/k} + kfn)$ for odd $k$  \cite{bodwin2021partial} & \multirow{2}{*}{$O(k(k+r)f^{1-1/(rk)}n^{1+1/k}\log n)$ (Ours)} \\
       & $O(k^2f^{1/2}n^{1+1/k} + kfn)$ for even $k$  \cite{bodwin2021partial} &   \\ \cline{2-3}
       & $\Omega(f^{1/2-1/(2k)}n^{1+1/k}+fn)$ \cite{bodwin2018optimal} & $\Omega(f^{1-\frac{1}{r}-\frac{1}{rk}+o(1)}n^{1+\frac{1}{k}-o(1)}+fn)$ (Ours) \\ \hline
      \multirow{2}{*}{VFT} & $O(f^{1-1/k}n^{1+1/k})$  \cite{bodwin2021optimal} &  Open   \\ \cline{2-3}
      & $\Omega(f^{1-1/k}n^{1+1/k})$ \cite{bodwin2018optimal} & $\Omega((f/r)^{r-1-1/k+o(1)}n^{1+1/k-o(1)})$ (Ours) \\ \hline
    \end{tabular}
    \caption{Size Bounds for $f$-Fault-Tolerant Spanners and Hyperspanners of stretch $2k-1$, where $n$ is the number of nodes, $r$ is the rank of the hypergraph. The lower bounds are conditional. We focus on odd stretch but even stretch is also very interesting to study.}
    \label{tab:results}
    \vspace{-0.2in}
\end{table*}

{\noindent \bf Contributions.}
In this paper, we initiate the study on fault-tolerant hyperspanners and develop the first edge fault-tolerant hyperspanners sublinear in $f$. We first provide a formal proof on applying the method of associated graphs for constructing non-faulty hyperspanners, carefully distinguishing the two settings of multi-graphs and simple graphs. We then show that two simple methods, the method of associated graphs and an adaptation of the peel-off method \cite{CLP+10}, produce size $O(fr^2n^{1+1/k})$ and $O(fn^{1+1/k})$ respectively, both linear in the number of allowed faults $f$. They still fall much behind the sublinear size bound in the graph counterpart \cite{bodwin2021partial}, as shown in the middle column of Table \ref{tab:results}.

As the major contribution, inspired by the FT clustering technique in \cite{partervft}, we propose a hypergraph clustering based algorithm and obtain a size $O(k(k+r)f^{1-1/(rk)}n^{1+1/k}\log n)$ sublinear in $f$ using fast runtime $\widetilde{O}(mr^3+nrf)$, as shown in Theorem \ref{thm:ub}. Furthermore, we accompany the upper bound with a lower bound $\Omega(f^{1-1/r-1/(rk)+o(1)}n^{1+1/k-o(1)}+fn)$ in Theorem \ref{thm:lb}, showing a separation from known graph bounds. We also provide a lower bound on the size of VFT hyperspanner $\Omega((f/r)^{r-1-1/k+o(1)}n^{1+1/k-o(1)})$ in Theorem \ref{thm:vft-lb}. 


\begin{theorem}[EFT Upper Bound]
\label{thm:ub}
    For an $n$-vertex $m$-hyperedge hypergraph of rank $r$ and a stretch parameter $k>1$, there is a randomized algorithm constructing an $f$-$EFT$ $(2k-1)$-hyperspanner of size $O(k(k+r)f^{1-1/(rk)}n^{1+1/k}\log n)$ in $\widetilde{O}(mr^3+nrf)$ time.
\end{theorem}

For the lower bounds, we need the conjecture of Spiro and Verstraëte, Conjecture I in \cite{spiro2022counting}:

\begin{conjecture}
\label{conjecture}
For all integers $l\geq 3$, $r\geq 2$ and $k=\lfloor{l/2}\rfloor$, the maximum number of hyperedges in $r$-uniform hypergraphs with girth at least $l+1$ is $n^{1+1/k-o(1)}$.
\end{conjecture}

\begin{theorem}[EFT Lower Bound]
\label{thm:lb}
    For integers $k\geq 1,f\geq1, r\geq 2$, assuming Conjecture \ref{conjecture}, there exist infinite families of undirected unweighted $n$-node $r$-uniform hypergraph with $\Omega(f^{1-1/r-1/(rk)+o(1)}n^{1+1/k-o(1)}+fn)$ hyperedges for which any $f$-$EFT$ $(2k-1)$-hyperspanner must contain all the hyperedges.
\end{theorem}

\begin{restatable}[VFT Lower Bound]{theorem}{thmvftlb}
\label{thm:vft-lb}
    For integers $k\geq 1,f\geq1, r\geq 2$, assuming Conjecture \ref{conjecture}, there exist infinite families of undirected unweighted $n$-node $r$-uniform hypergraph with $\Omega((f/r)^{r-1-1/k+o(1)}n^{1+1/k-o(1)})$ hyperedges for which any $f$-$VFT$ $(2k-1)$-hyperspanner must contain all the hyperedges.
\end{restatable}


{\noindent \bf Related Work.}
FT graph spanners have been subject to intensive study, such as \cite{CLP+10,bodwin2021optimal,bodwin2025lightedgefaulttolerant,petruschka2025color}, which push for optimal size bounds and stretch factors under different constraints. The question of optimal-sized VFT spanners has been settled, first in \cite{bodwin2018optimal} (for a fixed stretch) and then in \cite{bodwin19trivial}, achieving $O(f^{1-1/k}n^{1+1/k})$ for stretch $2k-1$. For EFT spanners, the best known lower bound on the size is $\Omega(f^{1/2-1/(2k)}n^{1+1/k}+fn)$ \cite{bodwin2018optimal}. For odd $k$, a matching upper bound was obtained, up to polynomial in $k$. For even $k$, the upper bound is $O(k^2f^{1/2}n^{1+1/k}+fkn)$, leaving a gap of $k^2f^{1/(2k)}$. Recent efforts have been devoted to optimal-time constructions of optimal-sized FT spanners \cite{bodwin2021optimal,dinitz2020efficient,partervft}. Dinitz and Robelle \cite{dinitz2020efficient} first developed a polynomial-time algorithm at the expense of a factor of $k$ in the size, which was further removed by Bodwin et al. \cite{bodwin2021optimal}. Optimal-time $\widetilde{O}(m)$ was achieved by Parter \cite{partervft} based on a novel FT clustering approach. Similar to \cite{partervft}, our method enjoys fast runtime and local computation, which is friendly to parallel and distributed computing. Unlike graph spanners, the literature on hyperspanners is scarce, with only a few known prior work \cite{oko2023nearly,pathak2024hypergraph,goranci2025fully}. \cite{oko2023nearly,goranci2025fully} adopted the associated graph idea to compute hyperspanners for spectral sparsfication while \cite{pathak2024hypergraph} empirically studied minimizing the number of hyperedges when mapping spanner edges of an associated graph to hyperedges.

\noindent \textbf{Preliminaries and notation.}
A weighted hypergraph is a triple $H=(V,E,w)$, where $V$ is a set of vertices, $E\subseteq 2^V$ is a set of hyperedges, and $w:E\to \mathbb{R}_{>0}$ assigns a positive weight to each hyperedge. The rank of $H$ is $r=\max_{h\in E}|h|$.
A (Berge) path from $s$ to $t$ in $H$ is an alternating sequence $s=v_0,h_1,v_1,h_2,\ldots,h_\ell,v_\ell=t$
such that $v_{i-1},v_i\in h_i$ for every $i\in[\ell]$. Its weight is $w(P)=\sum_{i=1}^{\ell}w(h_i)$.
The distance $\delta_H(u,v)$ is the minimum weight of a path from $u$ to $v$ in $H$, or $\infty$ if no such path exists. For an edge-fault set $F\subseteq E$, $H\setminus F$ denotes the hypergraph obtained by removing the hyperedges in $F$. For a vertex-fault set $F\subseteq V$, $H\setminus F$ denotes the hypergraph obtained by deleting the vertices in $F$ and all hyperedges incident to them. The girth of a hypergraph is the length of the shortest cycle.

\begin{definition}[EFT Hyperspanner]
    For a hypergraph $G = (V,E)$ and a subhypergraph $S(V, E' \subseteq E)$, $S$ is an $f$-edge fault tolerant (EFT) multiplicative $t$-hyperspanner if $\forall u, v \in V(G), \; \forall F \subseteq E(G), \; |F| \leq f$, 
    $\delta_{S \setminus F}(u, v) \leq t \cdot \delta_{G \setminus F}(u, v)$.
\end{definition}

\section{The Construction of Multiplicative EFT Hyperspanners}
\label{sec:multiplicative}

We start by discussing simple methods for constructing hyperspanners, and then present our main algorithm for the construction of sublinear-in-$f$ EFT hyperspanners.

\subsection{Simple Methods for EFT Hyperspanners}
Due to the page limit, we have to defer our discussion of the associated graph method to Appendix~\ref{apx:associated}. 
The peel-off method \cite{CLP+10} constructs $f$-EFT spanners in $f+1$ iterations. In each iteration, it computes the spanner of the current graph and then deletes the spanner edges from the current graph. Since the spanner edges in $f+1$ iterations are non-overlapping, their union contains $f+1$ parallel paths for every edge not included in the union, providing resilience to at most $f$ faulty edges. It is easy to see this method can be extended to construct hyperspanners and the proof is in Appendix~\ref{apx:peeloff}. Equipped with Lemma \ref{lem:hyperspanner}, this method produces a linear size $O(fn^{1+1/k})$ in the number of faults, still lagging behind the optimal VFT and partially optimal EFT spanners, both with a sublinearity in $f$.

Recent optimality results on FT spanners heavily use the greedy method \cite{bodwin2018optimal,dinitz2020efficient,bodwin2021partial}. Rooted at the greedy algorithm for optimal spanners \cite{ADD+93}, it checks for each edge whether there exists a fault set such that the stretch after the failure event does not hold. Many of these results use the blocking set to bound the size, which is a set of edge pairs such that for every cycle of small length, there must be an edge pair from the set on the cycle. The greedy method, if naively adapted to the hypergraph setting, has a technical barrier in bounding the size unfortunately. Specifically, for a small cycle, its last-added hyperedge may not be added due to the considered vertex pair on the cycle, but a different vertex pair for another cycle. It remains open to bound the output size of the greedy algorithm for constructing FT hyperspanners. See adaptations of other FT spanner methods in Appendix~\ref{apx:other}.

\subsection{Sublinear EFT Hyperspanners}

\noindent \textbf{High-level Overview.}
The goal of Section 2.2 is to obtain an EFT hyperspanner whose size is sublinear in the number of allowed hyperedge faults $f$. The associated-graph and peel-off approaches described above give natural baselines, but their size depends linearly on $f$. To improve this dependence, we adapt the fault-tolerant clustering framework of Parter \cite{partervft}. At a high level, this framework maintains, for each active vertex, many short paths to randomly sampled cluster centers. These paths play two roles. First, they provide fault tolerance since if sufficiently many of the stored paths are independent, then after at most $f$ failures at least one useful replacement path survives. Second, random sampling controls the size because as the levels progress, fewer centers remain active, so the algorithm can limit the number of paths that each vertex stores.

There are two main differences in the hypergraph setting. First, the failing objects are hyperedges. Thus, the relevant independence condition is hyperedge-disjointness of stored paths, rather than the graph condition used in the vertex-fault setting. Second, a hyperedge $h$ may contain many vertex pairs $u,v\in h$. The algorithm may have already certified a short replacement path for one pair in $h$ but not for another. For this reason, the algorithm assigns statuses to triples $(u,v,h)$ rather than only to hyperedges.

The algorithm proceeds in $k$ iterations. In iteration $i$, it has a set $Z_{i-1}$ of cluster centers, a set $V_{i-1}$ of active vertices, and for each active vertex $v$ a collection $Q_{i-1}(v)$ of paths from $v$ to centers in $Z_{i-1}$. It samples a smaller center set $Z_i$ and tries to extend existing paths by one hyperedge. When processing a triple $(u,v,h)$, the algorithm either keeps $h$ in the hyperspanner, safely discards the triple because enough short replacement paths already exist between $u$ and $v$, or postpones it to a later iteration. The stretch proof shows that every discarded triple has enough replacement paths to survive any $f$ hyperedge faults. The size proof shows that the sampling process prevents too many paths from being kept at any vertex.


\noindent \textbf{Details.}
A hyperedge differs from an ordinary edge in that it simultaneously connects many vertex pairs. Therefore, while processing a hyperedge $h$, the algorithm must reason separately about each pair $u,v\in h$. For some pairs, the current spanner may already contain enough short replacement paths; for others, the algorithm may still need to keep $h$ or postpone the decision. This is why the basic status object is the triple $(u,v,h)$. The status of $(u,v,h)$ records the algorithm's current decision about whether the hyperedge $h$ is needed to preserve the distance between $u$ and $v$. The status $\mathrm{kp}$ (keep) means that $h$ is kept in the hyperspanner. The status $\mathrm{sd}$ (safely discard) means that $h$ can be safely discarded for the pair $u,v$, because the current construction has already found sufficiently many short replacement paths between $u$ and $v$. The status $\mathrm{pp}$ (postpone) means that the algorithm has not yet certified either of these outcomes, so the decision is postponed to a later iteration.

Before giving the formal iteration, we summarize the state maintained by the algorithm. The set $Z_i$ contains the sampled centers at level $i$, and $V_i$ contains the vertices that remain active after level $i$. For each active vertex $v$, the collection $Q_i(v)$ stores the selected paths from $v$ to centers in $Z_i$ that will be used in the next iteration. During iteration $i$, the temporary collection $P_{i-1}(v)$ contains all paths currently stored for $v$, including paths that may be added to the hyperspanner but not necessarily passed to the next level. The set $R_i$ contains hyperedges that still have postponed pairs after iteration $i$, and $H_i$ is the partial hyperspanner constructed so far. Finally, $S_{i-1}(v)$ is a small random sample from $Q_{i-1}(v)$ used to test whether new candidate paths can be formed, and $K_f=12(k+r)f$ is the number of paths required for a vertex to remain active.



For a path $P$ from a center toward a vertex $v$, let $\operatorname{head}(P)$ denote the first hyperedge of $P$, namely the hyperedge containing the center endpoint of $P$. Let $\operatorname{head}(\emptyset)=\emptyset$. A path $P$ is passed to the next level only if $\operatorname{head}(P)$ contains a newly sampled center. Thus the head hyperedge determines whether $P$ survives under the next sampling step.

Alg.~\ref{alg:eft} is the main driver. In iteration $i$ (Alg.~\ref{alg:iteration}), we first randomly sample $Z_i$ from $Z_{i-1}$ with probability $p=f^{1/(rk)}/n^{1/k}$. We get a random sample $S_{i-1}(v)$ of $O(\log n)$ paths from $Q_{i-1}(v)$ (Line \ref{alg:sample}) and then for each vertex in $V_{i-1}$, process its hyperedges in $R_{i-1}$ in non-decreasing weight order (Line \ref{alg:h}). $P_{i-1}(v)$ is initialized to $Q_{i-1}(v)$. For each $(u,v,h)$ of status $pp$, if there exists a path $P \in S_{i-1}(u)$ whose hyperedges are disjoint with paths in $P_{i-1}(v)$ and whose head hyperedge is vertex-disjoint with head hyperedges of paths in $P_{i-1}(v)$, then we add the concatenation of $h$ and $P$ (denoted by $h \circ P$) to $P_{i-1}(v)$. 

The additional requirement that the head hyperedges be vertex-disjoint is used for the size analysis. A stored path survives to the next level when its head hyperedge contains a newly sampled center. If two stored paths had overlapping head hyperedges, then the same sampled center could influence both survival events. By maintaining vertex-disjoint head hyperedges, these sampling events are independent across the stored paths, which allows the Chernoff bound in Lemma~\ref{lem:bd} to be applied.

When adding a path $h \circ P$ to $P_{i-1}(v)$, we also check whether its head hyperedge $\operatorname{head}(P)$ contains a newly sampled center. If so, we add it to $Q_{i}(v)$. When examining $(u,v,h)$ of status $pp$, if we cannot find such an path, then we are safe to discard the triple (Line \ref{alg:sd}). We will prove that $h$ is not needed for vertex pair $u,v$. Finally, we include paths $P_{i-1}(v)$ for each $v\in V_{i-1}$ into $H_{i-1}$ to get $H_i$.

There are two sampling operations in the iteration. First, $S_{i-1}(v)$ is a small random sample of paths from $Q_{i-1}(v)$ used to test whether candidate concatenations exist. Second, after a new center set $Z_i$ is sampled, a path in $P_{i-1}(v)$ is called \emph{center-sampled} if its head hyperedge intersects $Z_i$. Only center-sampled paths are placed in $Q_i(v)$ and passed to the next level. This second sampling step is what controls the size of the path collections.

The sampling probability is chosen to balance the number of paths stored at each active vertex against the number of centers that survive to later levels. In the hypergraph setting, this balance depends on the rank $r$, because a sampled center is detected through a head hyperedge of size at most $r$, and the size analysis maintains vertex-disjointness among these head hyperedges. This leads to the choice $p=f^{1/(rk)}n^{-1/k}$,
which gives the target dependence $f^{1-1/(rk)}$ in the per-level path bound.


\begin{algorithm}[t]
\small
\caption{$f$-EFT $(2k{-}1)$-Hyperspanner}
\renewcommand{\algorithmicrequire}{\textbf{Input:}}
\renewcommand{\algorithmicensure}{\textbf{Output:}}
\begin{algorithmic}[1]
\Require $H(V,E)$
\Ensure $H'$

\State $Q_0 \gets \{\emptyset\,|\,v\in V\}$; $V_0 \gets V$; $R_0 \gets E$; $Z_0 \gets V$; $status \gets \{\}$; $H_0 \gets \emptyset$

\Statex
\State \textbf{Helper functions:}
\Function{getStatus}{$u,v,h$}
    \If{$status[h] = \textit{kp}$}
        \State \Return \textit{kp}
    \ElsIf{$(u,v) \in status[h]$ or $(v,u) \in status[h]$}
        \State \Return \textit{sd}
    \Else
        \State \Return \textit{pp}
    \EndIf
\EndFunction
\Statex
\Function{setStatusSd}{$u,v,h$}
    \State add $(u,v)$ to $status[h]$
\EndFunction
\Statex
\Function{setStatusKp}{$h$}
    \State $status[h] \gets \textit{kp}$
\EndFunction

\Statex

\For{$i \in [1,k]$}
    \State $(Q_i, V_i, R_i, Z_i, H_i) \gets \textsc{Clustering}(Q_{i-1}, V_{i-1}, R_{i-1}, Z_{i-1}, H_{i-1})$
\EndFor

\State \Return $H' \gets H_k$
\end{algorithmic}
\label{alg:eft}
\end{algorithm}

\begin{algorithm}[t]
\small
\caption{\textsc{Clustering}}
\renewcommand{\algorithmicrequire}{\textbf{Input:}}
\renewcommand{\algorithmicensure}{\textbf{Output:}}
\begin{algorithmic}[1]
\Require Collections of paths $Q_{i-1}$, Vertices $V_{i-1}$, Remaining hyperedges $R_{i-1}$, Cluster centers $Z_{i-1}$, and Hyperspanner $H_{i-1}$
\Ensure $Q_i, V_i, R_i, Z_i, H_i$

\If{$i \leq k - 1$}
    \State $Z_i \gets Z_{i-1}[f^{1/(rk)}/n^{1/k}]$ 
\Else
    \State $Z_i \gets \emptyset$
\EndIf

\ForAll{$v \in V_{i-1}$}
    \State $S_{i-1}(v) \gets$ a random sample of $O(\log n)$ paths from $Q_{i-1}(v)$  \label{alg:sample}
\EndFor

\ForAll{$v \in V_{i-1}$}
    \State $n_v \gets 0$
    \State $P_{i-1}(v) \gets Q_{i-1}(v)$
    \State $E_{i-1}(v) \gets \{h \in R_{i-1} \mid v \in h\}$ sorted by non-decreasing weight

    \ForAll{$h \in E_{i-1}(v)$} \label{alg:h}
        \ForAll{$u \in h$ \textbf{such that} \Call{getStatus}{$u, v, h$} = \textit{pp}}
            \If{there exists $P \in S_{i-1}(u)$ such that 
            $E(P) \cap E(P_{i-1}(v)) = \emptyset$ and 
            $V(head(P)) \cap V(head(P_{i-1}(v))) = \emptyset$} \label{alg:if}
                \State $P_{i-1}(v) \gets P_{i-1}(v) \cup (h \circ P)$ \label{alg:add}
                \State \Call{setStatusKp}{$h$}
                \If{$Z_i \cap \text{head}(P) \neq \emptyset$}
                    \State $n_v \gets n_v + 1$   \label{alg:nv}
                    \State $Q_{i}(v) \gets Q_{i}(v) \cup (h \circ P)$
                \EndIf
                \State \textbf{break}
            \Else
                \State \Call{setStatusSd}{$u, v, h$}  \label{alg:sd}
            \EndIf
        \EndFor
        \If{$n_v = K_f = 12(k + r)f$}
            \State \textbf{break}  \label{alg:stop}
        \EndIf
    \EndFor
\EndFor


\State $V_i \gets \{v \in V_{i-1} \mid |Q_i(v)| = K_f = 12(k + r)f\}$ \label{alg:Vi}
\State $Q_i \gets \{Q_i(v) \mid v \in V_i\}$
\State $H_i \gets \bigcup_{v \in V_{i-1}} P_{i-1}(v) \cup H_{i-1}$ \label{alg:union}
\State $R_i \gets \{h \notin H_i \mid$ \Call{getStatus}{$h$} $= \textit{pp}\}$
\State\Return $Q_i, V_i, R_i, Z_i, H_i$
\end{algorithmic}
\label{alg:iteration}
\end{algorithm}




We now analyze the algorithm in three steps. Lemma~\ref{lem:weight} records the main monotonicity and activity invariants maintained by the path collections. Lemmas~\ref{lem:half} and \ref{lem:paths} justify the $\mathrm{sd}$ status: if a triple is safely discarded, then the current hyperspanner already contains many short replacement paths for that vertex pair. Theorem~\ref{thm:stretch} uses these replacement paths to prove the stretch guarantee under any $f$ hyperedge faults. Finally, Lemma~\ref{lem:bd} and Theorem~\ref{thm:size} use the sampling of centers to bound the number of paths, and therefore the number of hyperedges, added to the hyperspanner.

\begin{lemma}
\label{lem:weight}
    (a) For vertex $v\in V_i$, the weight of hyperedges along the path $P\in Q_i(v)$ (towards $v$) are monotone non-decreasing.
    (b) For all $h\in R_i$ and $u,v\in h$, if $status_i(u,v,h)=pp$, then $u,v\in V_i$ and $w(h)>Q_i(u),Q_i(v)$.
\end{lemma}

\proof
We prove the two properties together by induction on iteration $i$.
When $i=0$, Property (a) holds clearly since $Q_0(v)=\emptyset$. Property (b) also holds because both $Q_0(v)$ and $Q_0(u)$ are empty. 
Assume both properties hold for iteration $i-1$. In iteration $i$, since $v\in V_i$, for any newly formed path $P\in P_{i-1}(v)$, by construction it is a concatenation of a path $P'\in Q_{i-1}(u)$ and a hyperedge $h\in R_{i-1}$ (Line \ref{alg:add}), where $u\in V_{i-1}$ and $status_{i-1}(u,v,h)=pp$. By the inductive hypothesis of Property (b), we know $w(h)>Q_{i-1}(u)$. By the inductive hypothesis of Property (a) on $u$, the weight of hyperedges along the path $P'$ is monotone non-decreasing. Therefore, the weight of hyperedges along the concatenation $h\circ P'=P$ is also monotone non-decreasing. Since $Q_{i}(v)\subseteq P_{i-1}(v)$, Property (a) holds for iteration $i$.

We now prove Property (b). Since $h\in R_i$, by definition, $h\not\in H_i$, then the if-test in Line \ref{alg:if} cannot be passed. The else-part of the if-test cannot be executed as well; otherwise, it contradicts $status_i(u,v,h)=pp$. Therefore, it must be that $h$ is not processed in the for-loop of Line \ref{alg:h} in both $u$ and $v$'s turn. Taking $v$ as an example, we have $|Q_i(v)|=K_f$ and $v\in V_i$. Because $E_{i-1}(v)$ is sorted by non-decreasing weight, $w(h)>w(LastEdge(P))$ for $P\in P_{i-1}(v)$. Since the weight of hyperedges for paths in $P_{i-1}(v)$ are monotone non-decreasing as proved, $LastEdge(P)$ is the heaviest hyperedge in the path $P$. Therefore, $w(h)>P_{i-1}(v)$, thus $w(h)>Q_i(v)$. By symmetry, we get $u,v\in V_i$ and $w(h)>Q_i(u),Q_i(v)$, completing the proof of Property (b). 
\qed

The next two lemmas explain why the algorithm is allowed to safely discard a triple. Suppose the algorithm examines $(u,v,h)$ but does not add $h$ through the pair $u,v$. Then the sampled tests suggest that many paths from $u$ already intersect, either through shared hyperedges or through overlapping head hyperedges, the paths currently stored for $v$. These intersections can be paired to form many short replacement paths between $u$ and $v$. Since there are more replacement paths than the number of possible faults, at least one survives any fault set of size at most $f$. Define $P_{i-1}(v,h)$ as the current $P_{i-1}(v)$ before examining $h$.

\begin{lemma} 
\label{lem:half}
    For a hyperedge $h\in R_{i-1}\backslash R_i, u,v\in h$ with $status_{i-1}(u,v,h)=pp$, if $h\not\in P_{i-1}(v)$, then w.h.p. at least a half of paths in $Q_{i-1}(u)$ either share a hyperedge with $P_{i-1}(v,h)$ or their head hyperedges share vertices with head hyperedges of $P_{i-1}(v,h)$.
\end{lemma}

\proof
Assume for contradiction that at least a half of paths in $Q_{i-1}(u)$ do not share a hyperedge with $P_{i-1}(v,h)$ and vertices in their head hyperedges do not intersect $w.h.p.$ Since we use uniform random sampling to sample $S_{i-1}(u)$ from $Q_{i-1}(u)$, we have at most $1/2$ probability to sample a path that either shares a hyperedge or its head hyperedge shares vertices with those of $P_{i-1}(v,h)$. Then, we have probability at most $1/2^{O(\log n)}=1/n^c$ (for $c>1$) that all the $O(\log n)$ sampled paths share a hyperedge or their head hyperedge share vertices with those of $P_{i-1}(v,h)$. Hence, $w.h.p.$ there exists a path $P\in S_{i-1}(u)$ that does not share a hyperedge and its head hyperedge does not share vertices with those of $P_{i-1}(v,h)$. In the algorithm, we would have added the path $h\circ P$ to $P_{i-1}(v)$, which violates the assumption that $h\notin P_{i-1}(v)$.
\qed


\begin{restatable}[]{lemma}{lempaths}
\label{lem:paths}
    For $h,u,v$ defined in Lemma \ref{lem:half}, there exist $2f+1$ distinct pairs of $(Q_j,P_j)\subset Q_{i-1}(u)\times P_{i-1}(v,h)$ such that $E(Q_j)\cap E(P_j)\neq\emptyset$ or $V(head(Q_j))\cap V(head(P_j))\neq\emptyset$ between u,v. The path formed by $Q_j,P_j$ has weight at most $(2i-1)w(h)$. All $Q_j$ are edge-disjoint and all $P_j$ are edge-disjoint.
\end{restatable}

{\noindent \em Proof Sketch.}
By Lemma \ref{lem:half}, we know there exist at least a half of paths in $Q_{i-1}(u)$ that share a hyperedge or their head hyperedges share vertices with those of $P_{i-1}(v,h)$. We want to find $2f+1$ distinct pairs of $(Q_j,P_j)\subset Q_{i-1}(u)\times P_{i-1}(v,h)$ such that $Q_j$ and $P_j$ share hyperedges or their head hyperedges share vertices, both of which result in an edge-disjoint path between $u$ and $v$ of weight at most $(2i-1)w(h)$. 

We will greedily pick $(Q_j,P_j)$ in $2f+1$ iterations. Let $B_1=P_{i-1}(v,h)$ and
\begin{equation*}
    \begin{aligned}
    A_1=\{Q\,|\,Q\in Q_{i-1}(u),E(Q)\cap E(P_{i-1}(v,h))\neq \emptyset\,\lor
    V(head(Q))\cap V(head(P_{i-1}(v,h)))\neq\emptyset\},
    \end{aligned}
\end{equation*}
where every path in $A_1$ intersects with some path in $B_1$. In iteration $j$, given $A_j,B_j$, first randomly pick $Q_j\in A_j$, and then randomly pick $P_j\in B_j$ that shares a hyperedge or its head hyperedge shares vertices with that of $Q_j$. Then update $B_{j+1}=B_j\backslash\{P_j\}$ and
\begin{equation*}
    \begin{aligned}
    A_{j+1}=\{Q\in A_j\,|\,E(P_j)\cap E(Q)=\emptyset\land V(head(P_j))\cap V(head(Q))=\emptyset\}
    \end{aligned}
\end{equation*}
To ensure we can find a valid pair in iteration $j$, we will show the invariants in the full proof: (1) $|A_j|>0$ and (2) for every $Q\in A_j$, $Q$ intersects with some path in $B_j$, i.e., $E(B_j)\cap E(Q)\neq\emptyset\lor V(head(B_j))\cap V(head(Q))\neq\emptyset$.
\qed

Before the stretch proof, we first note that $R_k=\emptyset$. Indeed, in the final iteration there is no subsequent center set to which a hyperedge can be postponed. Thus every triple $(u,v,h)$ that remains under consideration in the final iteration is either marked $\mathrm{kp}$, in which case $h$ is added to $H_k$, or marked $\mathrm{sd}$, in which case the pair $u,v$ has the replacement paths guaranteed by Lemma~\ref{lem:paths}. Therefore no hyperedge remains with status $\mathrm{pp}$ after iteration $k$, and hence $R_k=\emptyset$. With these, now we are ready to prove the stretch bound. 

\begin{theorem}[Stretch]
\label{thm:stretch}
    The spanner $H'$ returned by Algorithm \ref{alg:eft} is an $f$-$EFT$ ($2k-1$)-hyperspanner of $H$.
\end{theorem}

\proof
For vertices $u,v$ in $H$, let $\pi$ be their shortest path in $H\backslash F$. We show that there exists a path connecting them of weight at most $(2k-1)w(\pi)$ in $H'\backslash F$. For every hyperedge $h\in\pi$ missing from $H'$, we prove for every $x,y\in h$, we can find a path in $H'\backslash F$ of weight at most $(2k-1)w(h)$ between $x$ and $y$. To see this, since $R_0=E(H)$ and $R_k=\emptyset$, there must exist an iteration $j\in[k]$ such that $h\in R_{j-1}\backslash R_j$. By Lemma \ref{lem:paths}, we build $2f+1$ paths between $x$ and $y$.  Because all $Q_j$ are edge-disjoint and all $P_j$ are edge-disjoint, a hyperedge can appear in at most two paths. Given a fault set $F$ of size $|F|\leq f$, we must have at least one path that is not affected by $F$. Thus, we can always have a path between $x$ and $y$ of weight at most $(2i-1)w(h)$ in $H'\backslash F$. Therefore, we always have a path between $u$ and $v$ of weight $(2k-1)w(\pi)$ in $H'\backslash F$, completing the proof.
\qed


For the size bound, we will bound the number of hyperedges added in each iteration. Referring to Line \ref{alg:union} in Algorithm \ref{alg:iteration}, it depends on the size of $P_{i-1}(v)$. The following lemma helps achieve our size bound.

\begin{lemma}
\label{lem:bd}
    Fix a vertex $v\in V_{i-1}$ for $i<k$, and let $P_{i-1}(v)=\{P_1,...,P_l\}$ sorted in non-decreasing weights of their last hyperedges. If $|P_{i-1}(v)|>16(k+r)f^{1-1/(rk)}n^{1/k}\log n$, then w.h.p. the number of center-sampled paths in iteration $i$, $Y=|\{P\in P_{i-1}(v) \,|\, head(P)\cap Z_i\neq\emptyset\}|$ must have $Y> K_f$.
\end{lemma}

\proof
By construction, a path $P_j\in P_{i-1}(v)$ is sampled if its head hyperedge $h_j$ contains a center $z\in Z_i$. Let $Z^{h_j}_{i-1}=Z_{i-1}\cap h_j$ be the set of centers from $Z_{i-1}$ in $h_j$. Since $P_j\in P_{i-1}(v)$, we must have in iteration $i-1$, $|Z^{h_j}_{i-1}|\geq 1$. For the path $P_j$ to be sampled in iteration $i$, it must be $|Z^{h_j}_i|\geq 1$. Since each center is sampled independently with probability $p$, Pr$(|Z^{h_j}_i|\geq1)=1-Pr(|Z^{h_j}_i|=0)=1-(1-p)^{|Z^{h_j}_{i-1}|}$. Let $X_j$ be the indicator of whether $|Z^{h_j}_i|\geq1$ for each path $P_j\in P_{i-1}(v)$. Since head hyperedges of paths in $P_{i-1}(v)$ are vertex disjoint, the random variables $X_j$ for $1\leq j\leq l$ are i.i.d. Then we can use the Chernoff bound to analyze the probability of $Y=\sum X_j>K_f=12(k+r)f$.\\
\\
By definition, 
$$E[X_j]=Pr(|Z^{h_j}_i|\geq1)=1-(1-p)^{|Z^{h_j}_{i-1}|}\geq p.$$
Here the last inequality holds since $|Z^{h_j}_{i-1}|\geq 1$. It holds that
$$E[Y]=\sum_{j=1}^{|P_{i-1}(v)|} E[X_j]\geq |P_{i-1}(v)|\cdot p>16(k+r)f\cdot \log n,$$
where the last inequality follows since $|P_{i-1}(v)|>16(k+r)\cdot f^{1-1/(rk)}n^{1/k}\cdot \log n$ and $p=f^{1/(rk)}n^{-1/k}$.

By the Chernoff bound, Pr$(Y<(1-\epsilon)E[Y])\leq exp((-\epsilon^2/2)E[Y])=\frac{1}{n^{8(k+r)f\epsilon^2}}$. Because $k,r\geq 2,f\geq 1$, define $\epsilon \leq \frac{1}{4}$, then $\frac{1}{n^{8(k+r)f\epsilon^2}}\leq\frac{1}{n^2}$. Thus, $w.h.p$, $Y\geq \frac{3}{4}E[Y]=12(k+r)f\cdot \log n>K_f=12(k+r)f$.
\qed

\begin{theorem}[Size Bound]
\label{thm:size}
    The spanner $H'$ has size $O(k(k+r)f^{1-1/(rk)}n^{1+1/k}$ $\log n)$ $w.h.p.$
\end{theorem}

\proof
To bound the output $H'$ in the final iteration $k$, we focus on each iteration $i$ and bound the size of $H_i\backslash H_{i-1}$, i.e., $(\bigcup_{v\in V_{i-1}}P_{i-1}(v)) \backslash H_{i-1}$. Then we only need to bound $|P_{i-1}(v)|$. 

In our algorithm, we have the number of center-sampled paths $Y \leq K_f$ (i.e., $Y=K_f$ for vertices in $V_i$ and $Y<K_f$ for vertices not in $V_i$). According to the contrapositive of Lemma~\ref{lem:bd}, we get that $|P_{i-1}(v)|\leq 16(k+r)f^{1-1/(rk)}n^{1/k}\log n$ for $i<k$. In the last iteration $k$, we can bound $|P_{k-1}(v)|$ by using $|Z_{k-1}|$ because we are building paths from $v$ to $Z_{k-1}$ and head hyperedges of paths in $P_{k-1}(v)$ are vertex disjoint. Since w.h.p. $|Z_{k-1}|=O(np^{k-1}\log n)=O((k+r)f^{1-1/(rk)}n^{1+1/k}\log n)$, we get w.h.p. $|P_{k-1}(v)|\leq |Z_{k-1}|=O((k+r)f^{1-1/(rk)}n^{1+1/k}\log n)$. Then w.h.p. we add $O((k+r)f^{1-1/(rk)}n^{1+1/k}\log n)$ hyperedges in every iteration $i$. Therefore, by summing over $k$ iterations, w.h.p. the size is $O(k(k+r)f^{1-1/(rk)}n^{1+1/k}\log n)$.
\qed

The running time analysis is provided in Appendix~\ref{apx:runtime}.

\vspace{-0.2in}
\section{Lower Bounds}
Inspired by the lower bound technique of \cite{bodwin2018optimal}, we develop lower bounds on the sizes of $f$-$EFT$ and $f$-$VFT$ $(2k-1)$-hyperspanners. The idea is to create a new hypergraph $H'$ from the unweighted $r$-uniform hypergraph $H$ in Conjecture \ref{conjecture}, by producing $t$ copies $(u,1),(u,2),\cdots,(u,t)$ in $H'$ for each vertex $u$ in $H$, and adding hyperedge $\{(u_1,i_1),(u_2,i_2),\cdots,\\(u_r,i_r)\}$ to $H'$ for each hyperedge $\{u_1,u_2,\cdots,u_r\}$ in $H$ and $r$-tuple $i_1,i_2,\cdots,i_r\in [t]^r$. In the EFT and VFT settings, $t$ is set to $\lfloor f^{1/r} \rfloor$ and $\lceil f/r \rceil$, respectively. We then prove that $H'$ is the only $f$-EFT/VFT $(2k-1)$-hyperspanner of itself and analyze its size. The proof of the EFT lower bound (Theorem \ref{thm:lb}) is shown here and the proof of the VFT lower bound (Theorem \ref{thm:vft-lb}) is provided in Appendix~\ref{apx:missingproof}.

\proof
({\bf Theorem \ref{thm:lb}}) Consider an unweighted $r$-uniform hypergraph $H$ with girth at least $2k+2$. According to Conjecture \ref{conjecture}, it contains at least $n^{1+1/k-o(1)}$ hyperedges. We create a new hypergraph $H'$ with $V(H')=V(H)\times [t]$ for $t=\lfloor f^{1/r} \rfloor$, i.e., each vertex $u$ in $H$ has $t$ copies $(u,1),(u,2),\cdots,(u,t)$ in $H'$. Then for each hyperedge $\{u_1,u_2,\cdots,u_r\}\in H$ and $r$-tuple $i_1,i_2,\cdots,i_r\in [t]^r$, we create a hyperedge $\{(u_1,i_1),(u_2,i_2),\cdots,(u_r,i_r)\}$ in $H'$. That is, $E(H')=\{\{(u_1,$ $i_1),(u_2,i_2),\cdots,(u_r,i_r)\} \,|\, \{u_1,u_2,\cdots,u_r\}\in H \land i_1,i_2,\cdots,i_r\in [t]^r\}$.

We prove that $H'$ is the only $f$-EFT $(2k-1)$-hyperspanner of itself. Let $S$ be a subhypergraph of $H'$ missing some hyperedge $h=\{(u_1,i_1),(u_2,i_2)\cdots,(u_r,i_r)\}$. Let the fault hyperedge set $F$ be all hyperedges of $H'$ created for the hyperedge $\{u_1,u_2,\cdots,u_r\}\in H$ except the missing hyperedge $h$. Formally, $F=\{(u_1,i'_1),(u_2,i'_2)\cdots,(u_r,i'_r) \,|\, \exists j\in [r], i_j\not=i'_j\}$. Note that $h$ is the only hyperedge connecting any pair of its vertices $(u_j,i_j)$ and $(u_k,i_k)$ in $H'$, as otherwise there would be a cycle of length 2 in $H$, contradicting the girth of $H$. Let $P$ be the shortest path between $(u_j,i_j)$ and $(u_k,i_k)$ in $S\setminus F$ after the failure event. $P$ cannot contain $h$ since $h\not\in S$. $P$ cannot contain other hyperedges created for $\{u_1,u_2,\cdots,u_r\}$ since they are in the fault set $F$. Therefore, the distance of $P$ must be at least $2k+1$, because the girth is at least $2k+2$. In contrast, $h$ guarantees that the distance between $(u_j,i_j)$ and $(u_k,i_k)$ in $H\setminus F$ is 1. Hence, $S$ cannot be an $f$-EFT $(2k-1)$-hyperspanner of $H'$.

The number of hyperedges in $H'$ is 
\begin{equation*}
    \begin{aligned}
    |E(H')|&=t^r|E(H)|=\Omega(t^r|V(H)|^{1+1/k-o(1)})\\
    &=\Omega\left(t^r \left(\frac{|V(H')|}{t}\right)^{1+1/k-o(1)}\right)\\
    &=\Omega(t^{r-1-1/k+o(1)}n^{1+1/k-o(1)})\\
    &=\Omega(f^{1-1/r-1/(rk)+o(1)}n^{1+1/k-o(1)}).
    \end{aligned}
\end{equation*}
Here the second equality follows from Conjecture \ref{conjecture}. The $fn$ term is a simple folklore result stating that any f-regular hypergraph is the unique f-EFT spanner of itself.
\qed

When $r=2$ in the graph setting, our lower bounds degenerate to the lower bounds for EFT/VFT graph spanners \cite{bodwin2018optimal}. The bound $\Omega(f^{1-1/k}\cdot n^{1+1/k})$ on the size of VFT spanners is higher than the best known bound $\Omega(f^{1/2-1/(2k)}n^{1+1/k}+fn)$ for EFT spanners, demonstrating the inherent hardness of vertex fault tolerance. The lower bounds we derive for VFT and EFT hyperspanners, $\Omega((f/r)^{r-1-1/k+o(1)}n^{1+1/k-o(1)})$ and $\Omega(f^{1-1/r-1/(rk)+o(1)}n^{1+1/k-o(1)}+fn)$, further reveal the same observation in the hypergraph setting.

\balance                 
\bibliography{hyperspanner}

\appendix

\section{The Associated Graph Method}
\label{apx:associated}

For a hypergraph $H(V,F,z)$, the \emph{associated graph} (aka. clique expansion) $G(V,E,w)$ of $H$ is a \emph{multi-graph} obtained by replacing each hyperedge $h\in F$ with a clique $C(h)$ with $\binom{|h|}{2}$ edges of weights $z(h)$. A \emph{simple associated graph} $G’$ of $H$ is to remove all parallel edges but the \emph{lightest} edge (i.e., the edge of minimum weight with ties broken arbitrarily) between every pair of vertices in $H$’s associated graph $G$. Although the associated graph idea has been mentioned in some prior work \cite{oko2023nearly,goranci2025fully}, we have not seen a formal proof and more importantly, they do not distinguish multi-graphs and simple associated graphs, limiting its applicability.

\begin{restatable}[]{lemma}{lemassociated}
\label{lemma:associated}
Let $G$ be an associated graph of a hypergraph $H$ with a function $f:E(G)\rightarrow E(H)$ mapping each edge in $G$ to its hyperedge in $H$. If $G'$ is an $\alpha$-spanner of $G$ for $\alpha>1$, then the sub-hypergraph $H'$ of $H$ with the set of hyperedges $E(H')=\{f(e')\,|\,e'\in E(G')\}$ is an $\alpha$-hyperspanner of $H$. This also works when $G$ is a simple associated graph of $H$.
\end{restatable}

\proof
For vertices $u,v$ in $G$, by the definition of spanner, we have a path $P'$ in $G'$ of distance at most $\alpha\cdot d_G(u,v)$. Let $P$ be a path obtained by replacing each edge $e'$ in $P'$ with its corresponding hyperedge $f(e')$. It is easy to see that $P$ is a path between $u$ and $v$ in $H'$ of distance at most $\alpha\cdot d_G(u,v)=\alpha\cdot d_H(u,v)$. When $G$ is a simple associated graph, we still have $d_G(u,v)=d_H(u,v)$ since removed edges must not be on any shortest path in the original associated multi-graph. This completes the proof.

\qed


By using Lemma \ref{lemma:associated}, we can translate the results for any multiplicative spanners, whether they are for \emph{multi-graphs or simple graphs}, into corresponding hyperspanners of the same stretch. The size of the spanners also serves as an upper bound on the size of hyperspanners, although it is possible multiple spanner edges are mapped to one hyperedge, producing a smaller size. Then we have the following result for (non-faulty) hyperspanners. Note that it does not depend on the rank $r$, unlike all our FT results are parameterized on $r$.

\begin{restatable}[]{lemma}{lemhyperspanner}
\label{lem:hyperspanner}
Every hypergraph has a $(2k-1)$-hyperspanner of size $O(n^{1+1/k})$ for $k\geq 1$. The size is tight conditioned on the Erdös Girth conjecture.
\end{restatable}

\proof
By applying the algorithm of Althöfer et al. \cite{ADD+93} and Lemma \ref{lemma:associated}, we get the upper bound $O(n^{1+\frac{1}{k}})$.
Since graphs are a special case of hypergraphs, the lower bound $\Omega(n^{1+\frac{1}{k}})$ follows from the Erdös Girth Conjecture.
\qed

In addition to odd stretch, we can get the upper bounds on the size of hypergraph spanners of even sketch. By applying the Ruzsa-Szemeredi theorem \cite{ruzsa1978triple,conlon2021regularity}, which, in one form, states that a 3-regular hypergraph with $o(n^2)$ edges has girth at most 3, we get the upper bound $o(n^2)$ on 2-hyperspanners. Similarly, by applying the result that every $r$-regular hypergraph of girth greater than 5 has $o(n^{3/2})$ edges \cite{conlon2021regularity}, we get the upper bound $o(n^{3/2})$ on 4-hyperspanners. Note that we haven't provided formal proof and would encourage followup work to prove and claim these results.

It is conceivable to use the associated graph idea to construct EFT hyperspanners. Here we only consider associated multi-graphs since simple associated graphs combined with fault tolerance are not well-defined. But since each faulty hyperedge results in at most $r^2$ faulty edges in the associated graph, we have to set the number of edge faults to $fr^2$. Because the size of EFT spanners in multi-graphs was settled as $\Theta(fn^{1+1/k})$ \cite{petruschka2025color}, the size of EFT hyperspanners constructed by the associated graph method is $O(fr^2n^{1+1/k})$.

\section{Adaptation of the Peel-off Method \cite{CLP+10}}
\label{apx:peeloff}

\begin{algorithm}
\caption{Adapted Peel-off Algorithm \cite{CLP+10} for $f$-EFT $(2k-1)$-Hyperspanner}
\begin{algorithmic}[1]
\State \textbf{Input:} $H=(V,E)$
\State $E' \gets \emptyset$
\For{$i=1$ to $f+1$} 
    \State $(V,E'_i) \gets spanner(H\backslash E',k)$
    \State $E'=E'\cup E'_i$
\EndFor
\State \Return $H' \gets (V,E')$
\end{algorithmic}
\label{alg:peeloff}
\end{algorithm}

First, we prove the correctness of the algorithm. 
\begin{lemma}
    Algorithm \ref{alg:peeloff} returns an $f$-EFT $k$-hyperspanner $H'$ of $H$.
\end{lemma}

\begin{proof}
    For an arbitrary $F\subseteq E(H),|F|\leq f$, for any pair of vertices $u,v\in V(H)$, let $\pi$ be the minimum path of $u,v$ in $H\backslash F$. Hence, we want to show that for any hyperedge and $h_i\in\pi$, for any $u_i,v_i\in h_i,dist_{H'\backslash F}(u_i,v_i)<(2k-1)w(h_i)$.
    
    If $h_i\in E(H')\backslash F$, then we are good because the stretch threshold trivially holds. If not, it means that for any $u_i,v_i\in h_i$, we can find $f+1$ alternative path such that the length is smaller than $(2k-1)w(h_i)$. Specifically, it is possible that $u_i,v_i\in h^0_i,..,h^n_i$ where $w(h^0_i)\leq w(h^1_i)\leq \dots \leq w(h^n_i)$ and assume $h^i_i=h_i$, but we could simply treat each hyperedge as a $u_i,v_i$ path. Notice that spanner in each iteration are edge disjoint. If for all $f+1$ iterations, $w(h_i)$ does not appear as the minimum $u_i,v_i$ path length, then in each iteration $j$ we have a $u_i,v_i$ path that satisfy $dist_{H'\backslash E'}(u_i,v_i)<(2k-1)\times dist_{H\backslash E'}(u_i,v_i)<(2k-1)w(h_i)$. And because $h_i$ is not included in the spanner, there's no concern that we skip the comparison to $(2k-1)w(h_i)$ as $h_i\in H\backslash E'$. And if it appears in iteration $j$, then in the previous iterations the same argument hold that we can find $j-1$ $u_i,v_i$-path within the threshold. And from iteration $j$, since $h_i$ is not included in the spanner, it means that we still can find $u_i,v_i$-path satisfy $dist_{H'\backslash E'}(u_i,v_i)<w(h_i)$ until the end.
    
    Since $|F|\leq f,H'\backslash F$ can fail at most $f$ $u_i,v_i-$path, which means we always have one path unaffected and within the threshold. Therefore, $dist_{H'\backslash F}(u,v)=\sum_{(u_i,v_i)\in h_i\in \pi} dist(u_i,v_i)< \sum_{h_i\in \pi} (2k-1)w(h_i)=(2k-1)\times dist_{H\backslash F}(u,v)$.
\end{proof}

Then, we show the size of the spanner.
\begin{lemma}
    The spanner size returned by Algorithm \ref{alg:peeloff} is $O(fn^{1+\frac{1}{k}})$.
\end{lemma}

\begin{proof}
    Lemma \ref{lem:hyperspanner} states that there exists a (2k-1)-hyperspanner of size $O(n^{1+1/k})$. Since the algorithm takes the union of $f$ such hyperspanner, the output has size $O(fn^{1+1/k})$.
    
\end{proof}

\section{Adaptions of other FT spanner methods to FT hyperspanners}
\label{apx:other}

A hyperedge may also be viewed as an edge color in the clique expansion: all clique edges generated by the same hyperedge receive the same color. Applying known edge-color-fault-tolerant spanner bounds \cite{petruschka2025color} and mapping graph edges back to their generating hyperedges yields $O(fn^{1+1/k})$ hyperedges, matching the linear dependence of the peel-off baseline but not the sublinear dependence sought here. Alternatively, applying a weighted $f$-VFT spanner construction to the ($n + m$)-vertex incidence graph yields $O(f^{1-1/k}(n + m)^{1+1/k})$ selected hyperedges. In specific, form the incidence graph $B_H = (V \cup E(H), I)$ with $I$ indicating the incidence between nodes on one side and hyperedges on the other side. For weighted hypergraphs, give each incidence edge adjacent to hyperedge $h$ weight $w(h)/2$. A hypergraph path and its corresponding path between original vertices in $B_H$ then have the same weight. A hyperedge failure becomes a vertex failure on the $E(H)$-side of $B_H$. Applying a
weighted $f$-VFT graph-spanner bound gives $O(f^{1-1/k}(n + m)^{1+1/k})$ graph edges and hence at most that many selected hyperedges after mapping back.
This is useful when $m$ is small, but does not provide a bound in terms of $n$ alone for general hypergraphs. Thus neither reduction subsumes our claimed n-based sublinear-in-$f$ construction.

Alternatively, one can subsample the nodes of the $m$-node side with probability $\sim 1/f$, yielding a bipartite graph with $n$ nodes on one side, $m/f$ nodes on the other side, and high girth. Then we would apply the extremal bounds for such graphs, and multiply by $f$ (to account for the $\sim 1/f$ fraction of the edges that did not survive the subsample). This leads to $O(f(n+m/f)^{1+1/k})$ hyperedges.

\section{Missing Proof}
\label{apx:missingproof}

\lempaths*

\proof
({\bf Lemma \ref{lem:paths}}) 
By Lemma \ref{lem:half}, we know there exist at least a half of paths in $Q_{i-1}(u)$ that share a hyperedge or their head hyperedges share vertices with those of $P_{i-1}(v,h)$. We want to find $2f+1$ distinct pairs of $(Q_j,P_j)\subset Q_{i-1}(u)\times P_{i-1}(v,h)$ such that $E(Q_j)\cap E(P_j)\neq\emptyset$ or $V(head(Q_j))\cap V(head(P_j))\neq\emptyset$.

We will greedily pick $(Q_j,P_j)$ in $2f+1$ iterations. Let 
\begin{equation*}
    \begin{aligned}
    A_1=\{Q\,|\,Q\in Q_{i-1}(u),E(Q)\cap E(P_{i-1}(v,h))\neq \emptyset\,\lor
    V(head(Q))\cap V(head(P_{i-1}(v,h)))\neq\emptyset\}
    \end{aligned}
\end{equation*}
and $B_1=P_{i-1}(v,h)$. Note that $|A_1|\geq K_f/2=6(k+r)f$ by Lemma \ref{lem:half}.  In iteration $j$, given $A_j,B_j$, first randomly pick $Q_j\in A_j$, and then randomly pick $P_j\in B_j$ that shares a hyperedge or its head hyperedge shares vertices with that of $Q_j$. Then update $B_{j+1}=B_j\backslash\{P_j\}$ and
\begin{equation*}
    \begin{aligned}
    A_{j+1}=\{Q\in A_j\,|\,E(P_j)\cap E(Q)=\emptyset\land V(head(P_j))\cap V(head(Q))=\emptyset\}
    \end{aligned}
\end{equation*}
To ensure we can find a valid pair in iteration $j$, we need to show (1) $|A_j|>0$ and (2) for every $Q\in A_j$, $E(B_j)\cap E(Q)\neq\emptyset\lor V(head(B_j))\cap V(head(Q))\neq\emptyset$.

Since $|P_j|\leq k$ for $P_j\in P_{i-1}(v,h)$, $P_j$ can share at most $k$ hyperedges with $A_1$. Because paths in $A_1$ are edge-disjoint, $P_j$ can share hyperedges with at most $k$ paths in $A_1$. Since the head hyperedge of $P_j$ contains at most $r$ vertices and all head hyperedges in $A_1$ are vertex-disjoint, $P_j$ can share vertices in head hyperedges with at most $r$ paths in $A_1$. Thus, $P_j$ can share hyperedges or share vertices in head hyperedges with at most $k+r$ paths in $A_1$. Therefore, $|A_{j}|\geq |A_1|-(k+r)j\geq 6(k+r)f-(k+r)(2f-1)=(k+r)(4f-1)>0$. 

We prove (2) by induction. The base case holds by the definition of $A_1$ and $B_1$. Assume (2) holds for $j$. Let $P_j$ be the path we delete in $B_j$. By the way we maintain $A_{j+1}$ and $B_{j+1}$, it is easy to see that $\forall Q\in A_{j+1}$, $E(B_{j+1})\cap E(Q)\neq\emptyset\lor V(head(B_{j+1}))\cap V(head(Q))\neq\emptyset$.

Since all paths in $Q_i(u)$ are edge-disjoint and all paths in $P_i(v,h)$ are edge-disjoint, then all $Q_j$ are edge-disjoint and all $P_j$ are edge-disjoint. Because $status_{i-1}(u,v,h)=pp$, by Lemma \ref{lem:weight}(b), we have $w(h)>Q_{i-1}(u)$. By the fact that $P_{i-1}(v,h)$ is $P_{i-1}(v)$ before processing $h$ and the monotone non-decreasing property in Lemma \ref{lem:weight}(a), we get $w(h)>LastEdge(P_{i-1}(v,h))\geq P_{i-1}(v,h)$. Because $|Q_j|\leq i-1$ and $|P_j|\leq i$, it holds that $w(Q_j)+w(P_j)\leq (2i-1)w(h)$. Therefore, there exist $2f+1$ paths between $u,v$ of weight at most $(2i-1)w(h)$.
\qed

\thmvftlb*

\proof
({\bf Theorem \ref{thm:vft-lb}}) Consider an unweighted $r$-uniform hypergraph $H$ with girth at least $2k+2$. According to Conjecture \ref{conjecture}, it contains at least $n^{1+1/k-o(1)}$ hyperedges. We create a new hypergraph $H'$ with $V(H')=V(H)\times [t]$ for $t=\lceil f/r \rceil$, i.e., each vertex $u$ in $H$ has $t$ copies $(u,1),(u,2),\cdots,(u,t)$ in $H'$. Then for each hyperedge $\{u_1,u_2,\cdots,u_r\}\in H$ and $r$-tuple $i_1,i_2,\cdots,i_r\in [t]^r$, we create a hyperedge $\{(u_1,i_1),(u_2,i_2),\cdots,(u_r,i_r)\}$. That is, $E(H')=\{(u_1,$ $i_1),(u_2,i_2),\cdots,(u_r,i_r)\} \,|\, \{u_1,u_2,\cdots,u_r\}\in H \land i_1,i_2,\cdots,i_r\in [t]^r\}$.

We prove that $H'$ is the only $f$-VFT $(2k-1)$-hyperspanner of itself. Let $S$ be a subgraph of $H'$ missing some hyperedge $h=\{(u_1,i_1),(u_2,i_2)\cdots,(u_r,i_r)\}$. We then let the fault vertex set $F$ be all vertices created for $u_1,u_2,\cdots,u_r$ except for those in $h$. Formally, $F=\{(u_1,l)\,|\, l\not= i_1\} \cup \{(u_2,l)\,|\, l\not= i_2\}\cup \cdots \cup \{(u_r,l)\,|\, l\not= i_r\}$. Note that $h$ is the only hyperedge connecting any of its vertex pairs $(u_j,i_j)$ and $(u_k,i_k)$ in $H'$, as otherwise there would be a cycle of length 2 in $H$, contradicting the girth of $H$. Let $P$ be the shortest path between $(u_j,i_j)$ and $(u_k,i_k)$ in $S\setminus F$ after the failure event. $P$ cannot contain $h$ since $h\not\in S$. $P$ also cannot contain other hyperedges created for $\{u_1,u_2,\cdots,u_r\}$ since their endpoints are in the fault set $F$. Therefore, the distance of $P$ must be at least $2k+1$, because the girth is at least $2k+2$. In contrast, $h$ guarantees that the distance between $(u_j,i_j)$ and $(u_k,i_k)$ in $H\setminus F$ is 1. Hence, $S$ cannot be an $f$-VFT $(2k-1)$-hyperspanner of $H'$ and the only $f$-VFT $(2k-1)$-hyperspanner of $H'$ is itself.

The size of $H'$ is 
\begin{equation*}
    \begin{aligned}
    |E(H')|&=t^r|E(H)|=\Omega(t^r|V(H)|^{1+1/k-o(1)})\\
    &=\Omega\left(t^r \left(\frac{|V(H')}{t}\right)^{1+1/k-o(1)}\right)\\
    &=\Omega(t^{r-1-1/k+o(1)}n^{1+1/k-o(1)})\\
    &=\Omega((\frac{f}{r})^{r-1-1/k+o(1)}n^{1+1/k-o(1)}).
    \end{aligned}
\end{equation*}
Here the second equality follows from Conjecture \ref{conjecture}.
\qed

\section{Running Time Analysis}
\label{apx:runtime}
We show that Algorithm \ref{alg:iteration} has runtime $\widetilde{O}(mr^3+nrf)$. First, sampling cluster centers takes $(|Z_{i}|)=\widetilde{O}(n)$ time, and sampling paths $S_{i-1}$ takes the same time. For each vertex $v$, we loop through its associated hyperedges $h$ in the outer for-loop and each vertex $u$ of $h$ (with status $pp$) in the inner for-loop, which result in $deg(v)\cdot r$ iterations. Then the total number of iterations for all vertices is $mr^2$. When storing the status of all triples in $O(mr^2)$ space, checking and setting statuses only incur constant time. In each iteration, we check each of the $O(\log n)$ paths in $S_{i-1}(u)$ whether its hyperedges intersect with $P_{i-1}(v)$ and its head hyperedge shares vertices with those of $P_{i-1}(v)$. Since each path has at most $k$ hyperedges and a hyperedge has size at most $r$, this incurs $O(k+r)$ time. Similarly, checking the intersection of the head hyperedge and $Z_i$ incurs $O(r)$ time. In addition, we process each vertex and construct $O((k+r)f)$ paths in $Q_i$, which takes $O(n(k+r)f)$ time. Therefore, Algorithm \ref{alg:iteration} incurs $\widetilde{O}(mr^2(k+r)+n(k+r)f)$ time. Summing over $k$ iterations, the construction of EFT hyperspanners (Alg.~\ref{alg:eft}) has runtime $\widetilde{O}(mr^2k(k+r)+nk(k+r)f)$. Since $k$ is typically $O(\log n)$, the runtime can be simplified as $\widetilde{O}(mr^3+nrf)$.

\end{document}